\documentclass[aps,pre
,tightenlines
,twocolumn
,showpacs
,amsfonts,amsmath,amssymb
]{revtex4}

\usepackage{graphicx}

\begin{document}

\title{Bubble-bubble interaction: A potential source of cavitation noise}

\author{Masato Ida}
\email[E-Mail: ]{ida.masato@jaea.go.jp}
\altaffiliation[Present address: ]{Materials and Life Science Division, J-PARC Center, Japan Atomic Energy Agency, 2-4 Shirakata-Shirane, Tokai-mura, Naka-Gun, Ibaraki 319-1195, Japan}
\affiliation{Center for Computational Science and E-systems, Japan Atomic Energy Agency, Higashi-Ueno, Taito-ku, Tokyo 110-0015, Japan}

\begin{abstract}
The interaction between microbubbles through pressure pulses has been 
studied to show that it can be a source of cavitation noise. A recent report 
demonstrated that the acoustic noise generated by a shrimp originates from 
the collapse of a cavitation bubble produced when the shrimp closes its 
snapper claw. The recorded acoustic signal contains a broadband noise that 
consists of positive and negative pulses, but a theoretical model for single 
bubbles fails to reproduce the negative ones. Using a nonlinear multibubble 
model we have shown here that the negative pulses can be explained by 
considering the interaction of microbubbles formed after the cavitation 
bubble has collapsed and fragmented: Positive pulses produced at the 
collapse of the microbubbles hit and impulsively compress neighboring 
microbubbles to generate reflected pulses whose amplitudes are negative. 
Discussing the details of the noise generation process, we have found that 
no negative pulses are generated if the internal pressure of the reflecting 
bubble is very high when hit by a positive pulse.
\end{abstract}


\maketitle

\section{Introduction}

In a recent paper, Versluis {\it et al}.~reported that the snapping shrimp ({\it Alpheus heterochaelis}) living in 
the ocean can generate a cavitation bubble by rapidly closing its large 
snapper claw \cite{ref1}. The rapid closure produces a negative pressure in 
seawater, by which cavitation nuclei (e.g., air microbubbles) are 
explosively expanded to a radius of a few millimeters. The cavitation bubble 
then collapses violently and emits a loud acoustic noise. The experimentally 
recorded acoustic signal presented in Ref.~\cite{ref1} consists of a strong 
(positive) pressure pulse, clearly produced at the bubble collapse, and a 
subsequent broadband noise \cite{ref16}. A single-bubble theoretical model (the 
Keller-Miksis equation) succeeded to reproduce the strong pressure pulse 
(and a weak precursor signal) but failed to describe the broadband noise. 
Versluis {\it et al}.~stated that the broadband noise is, partly, due to the 
reflection of the pressure pulse at nearby aquarium walls. However, the 
broadband noise begins earlier than the reflected wave reaches, immediately 
after the bubble collapse.

The broadband noise appears to consist of both positive and negative 
pressure pulses (or steep spikes) whose amplitudes are smaller than the 
first strong pulse. As demonstrated in Ref.~\cite{ref1} and shown below, however, 
the single-bubble model cannot describe negative pulses. A key to resolving 
this inconsistency is given from an image recording of the cavitation 
bubble. In a series of high-speed images, it was found that at collapse the 
single cavitation bubble breaks apart through the surface instability and 
then an opaque cloud of microbubbles appears \cite{ref1}. The bubble cloud seems to 
grow and finally dissolve away.

We hypothesize that the interaction between the microbubbles through 
pressure pulses is a source of the negative pulses involved in the broadband 
noise. It is known that bubbles undergoing volume change interact with each 
other through the pressure waves that they emit. Bubble-bubble interaction 
of this type leads to a variety of phenomena, including attraction/repulsion 
between bubbles \cite{ref2,ref3,ref4,refRefadd7,ref5,ref6}, filamentary structure formation 
\cite{refRefadd8}, change of eigenfrequencies \cite{ref7}, superresonances \cite{refRefadd9}, 
emergence of transition frequencies \cite{ref6,ref8}, avoided crossing of resonance 
frequencies \cite{ref9}, sound localization \cite{ref10}, and suppression of cavitation 
inception \cite{ref11,ref12}. Because at generation the microbubbles are highly 
compressed, they must begin volume change (as observed) and emit pressure 
pulses at their collapse. Taking the bubble-bubble interaction through the 
pressure pulses into consideration, in this paper we suggest a possible 
origin of the negative pulses. Here we do not aim at providing a 
quantitative explanation because the actual sizes and number density of the 
microbubbles are now unknown. We instead attempt to elucidate the basic 
mechanism of negative pulse generation, reducing the problem to the 
interaction of only two bubbles.

The rest of this paper is organized as follows: In Sec.~\ref{secII}, the model 
equations and assumptions used in this study are introduced. 
Section \ref{secIII} presents numerical and theoretical results and discussions 
on the pressure pulses emitted by interacting bubbles, and Sec.~\ref{secIV} 
summarizes the obtained results.

\section{Model equations}
\label{secII}

The theoretical model used in this study is the coupled Keller-Miksis 
equations \cite{ref4,ref12}, which describe the radial motion of two coupled spherical 
bubbles in a compressible liquid:
\begin{eqnarray}
\label{eq1}
&& \left( {1-\frac{\dot {R}_i }{c}} \right)R_i \ddot {R}_i +\left( 
{\frac{3}{2}-\frac{\dot {R}_i }{2c}} \right)\dot {R}_i^2 =\frac{1}{\rho 
}\left( {1+\frac{\dot {R}_i }{c}} \right)p_{s,i} \nonumber \\ 
&& \quad \quad \quad \quad +\frac{R_i }{\rho c}\frac{d}{dt}p_{s,i} 
-\sum\limits_{j=1,j\ne i}^2 {\frac{1}{D_{ij} }\frac{d(R_j^2 \dot {R}_j 
)}{dt}} ,
\end{eqnarray}
\begin{equation}
\label{eq2}
p_{s,i} \equiv p_{b,i} -\frac{2\sigma }{R_i }-\frac{4\mu \dot {R}_i }{R_i 
}-P_0 ,
\end{equation}
\[
\mbox{for}\quad i=1,2,
\]
where $R_i =R_i (t)$ is the instantaneous radius of bubble $i$, $D_{ij} $ is 
the center-to-center distance between bubbles $i$ and $j$, and the overdots 
denote the time derivative $d/dt$. The surrounding liquid is assumed to be 
water with density $\rho =1000$ kg/m$^{3}$, viscosity $\mu =1.002\times 
10^{-3}$ kg/(m s), sound speed $c=1500$ m/s, and surface tension $\sigma 
=0.0728$ N/m. The far-field pressure $P_0 $ is assumed to be constant in 
time and equal to the atmospheric pressure, $0.1013$ MPa. The gas in the 
bubbles is assumed to be a van der Waals gas (air) and the pressure inside 
bubble $i$ ($p_{b,i} )$ is determined by
\begin{equation}
\label{eq3}
p_{b,i} =\left( {P_0 +\frac{2\sigma }{R_{i0} }} \right)\left( 
{\frac{R_{i0}^3 -h_i^3 }{R_i^3 -h_i^3 }} \right)^\kappa 
\end{equation}
where $R_{i0} $ is the ambient radius and $h_i $ is the hard-core radius 
($R_{i0} /8.54$ for air \cite{ref13}). The polytropic exponent of the gas, $\kappa 
$, is assumed to be equal to its specific heat ratio $\gamma $ (1.4 for air) 
because our interest is in the pressure pulses emitted at bubble collapse, 
where the bubble behavior is nearly adiabatic. We confirmed numerically that 
even when the heat exchange between the bubbles and water is taken into 
account by, e.g., a switching method for $\kappa $ \cite{refRefadd1}, results are 
essentially the same as those shown below. The vapor pressure is neglected 
for simplicity. We do not consider mass exchange (i.e., 
evaporation/condensation, mass diffusion) and chemical reactions, which may 
occur inside the bubbles \cite{ref14}, since they are not essential for sound 
emission: As demonstrated in earlier work on single bubbles \cite{refRefadd2,refRefadd3}, models that do not take mass exchange and chemical reactions into 
account can describe bubble emitted pressure waves with sufficient accuracy.

The last term of Eq.~(\ref{eq1}) describes the bubble-bubble interaction through the 
bubble emitted pressure waves and acts as a driving force on bubble $i$. 
This term was derived from the following simple formula, which corresponds 
to an equation for the pressure wave emitted by a pulsating sphere,
\begin{equation}
\label{eq4}
p_i =\frac{\rho }{r_i }\frac{d(R_i^2 \dot {R}_i )}{dt}=\frac{\rho }{r_i 
}(2R_i \dot {R}_i^2 +R_i^2 \ddot {R}_i ),
\end{equation}
where $r_i $ is the distance measured from the center of bubble $i$. This 
pressure equation can be given from the continuity and Euler equations of 
fluid flow (see, e.g., Refs.~\cite{ref4,ref12}), and is in the following used to 
examine the acoustic signal from the bubbles. Time-delay effects \cite{refRefadd4} 
due to the finite sound speed of water are neglected, but a remark will be 
made on a consequence of it. As a first approximation we neglect the 
translational motion of bubbles due to the secondary Bjerknes force (an 
interaction force proportional to $D_{ij}^{-2} )$ \cite{refRefadd6}, that is, 
assuming $dD_{ij} /dt=0$. Hence, we set $D_{ij} $ to be much larger than 
$R_{i0} $ ($D_{ij} \approx 10R_{i0} )$ and consider only the first few 
periods of bubble oscillation, in which the translational velocity should be 
very small. As demonstrated below, the first one or two periods are 
sufficient for our discussion.

\section{Results and discussions}
\label{secIII}

\begin{figure} 
\includegraphics[width=7.8cm]{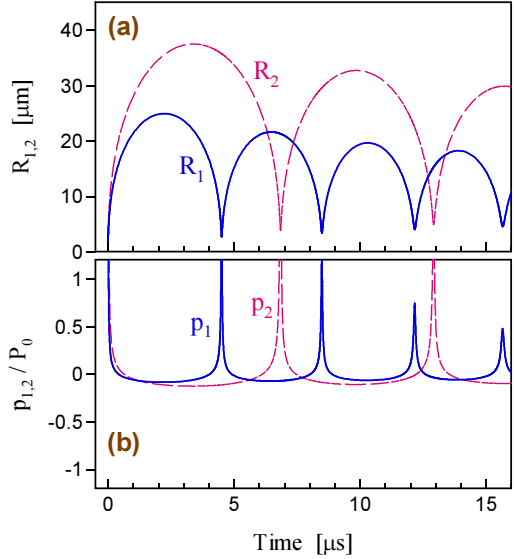}
\caption{(Color online) Bubble radii $R_{1,2} $ (a) and emitted pressures 
$p_{1,2} $ measured at $r_{1,2} =30R_{10} $ (b) in a single-bubble case 
($D_{12} =\infty )$ as functions of time. The ambient radii of the bubbles are 
$R_{10} =10$ $\mu $m and $R_{20} =15$ $\mu $m. The solid and dashed lines are 
for bubbles 1 and 2, respectively. The pressures are normalized by the 
atmospheric pressure $P_0 $. Only positive pulses are found in $p_{1,2} $.}
\label{fig1}
\end{figure}

First we consider a single-bubble case to confirm that single bubbles can 
emit positive pulses only. The radius-time curves for two different bubbles 
of $R_{10} =10$ $\mu $m, $R_{20} =15$ $\mu $m with $D_{12} =\infty $ are shown 
in Fig.~\ref{fig1}(a). Here, we set the initial condition as $R_i 
(t=0)=\mbox{0.13}R_{i0} $ and $\dot {R}_i (t=0)=0$ in order to simulate the 
highly compressed state of the microbubbles at their generation. Due to the 
high internal pressure, the bubbles undergo rapid expansion and reach a 
maximum size of about $2.5R_{i0} $. Then they collapse and rebound many 
times. The oscillation amplitudes decrease monotonically in time due to the 
viscosity and compressibility of water. The pressure waves from the bubbles 
measured at $r_1 =r_2 =30R_{10} $ are shown in Fig.~\ref{fig1}(b). From this, one 
knows that the single bubbles can emit positive pulses only. Between the 
positive pulses, one finds low-amplitude negative pressures, which resulted 
from a weak deceleration of the bubble surface acting when $R_i (t)>R_{i0} 
$. Their amplitudes (and also fundamental frequencies) are, however, 
obviously much lower than those of the positive pulses and hence they are 
incapable of explaining the negative pulses in the acoustic signal. The 
positive pulses are produced at the bubble collapse where the bubble surface 
is strongly accelerated [i.e., $\ddot {R}_i $ in Eq.~(\ref{eq4}) has a very large 
value]. However, from Eq.~(\ref{eq4}) one finds that to produce a negative pulse the 
bubble surface needs to be strongly decelerated: this may be impossible when 
only single bubbles are considered.

\begin{figure} 
\includegraphics[width=7.8cm]{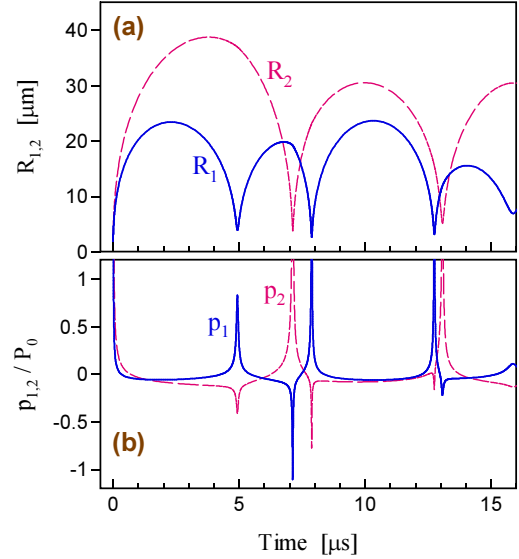}
\caption{(Color online) Same as Fig.~\ref{fig1}, but in a double-bubble case ($D_{12} 
=10R_{10} )$. Both positive and negative pulses are found in $p_{1,2} $, as 
in the broadband noise reported in Ref.~\cite{ref1}.}
\label{fig2}
\end{figure}

\begin{figure} 
\includegraphics[width=7.8cm]{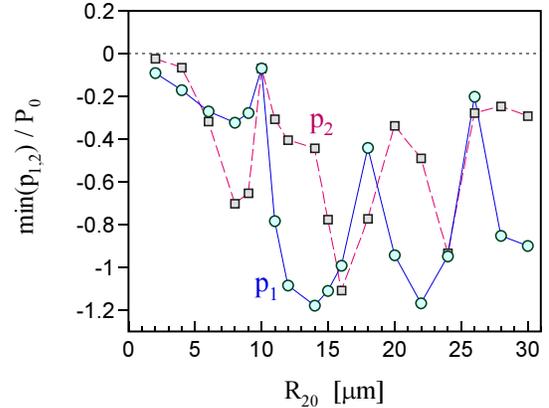}
\caption{(Color online) Maximum amplitudes of the negative pulses normalized 
by $P_0 $ [$\min (p_{1,2} )/P_0 $] as functions of $R_{20} $. $R_{10} $ and 
$D_{12} $ are fixed to $10$ $\mu $m and $10R_{10} $, respectively. The circles 
and squares are for bubbles 1 and 2, respectively.}
\label{fig3}
\end{figure}

\begin{figure} 
\includegraphics[width=7.8cm]{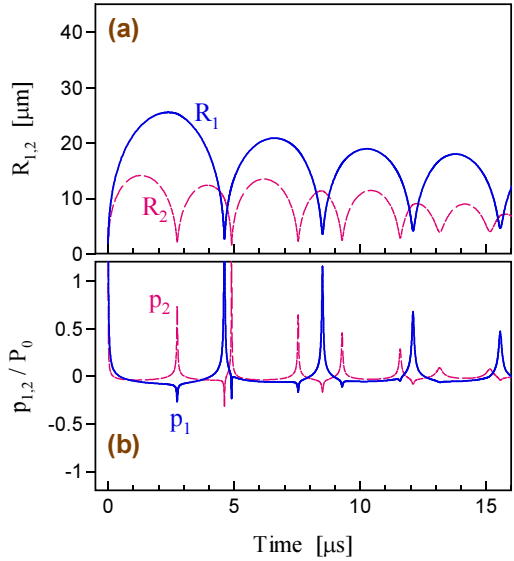}
\caption{(Color online) Same as Fig.~\ref{fig2}, but for $R_{10} =10$ $\mu $m and 
$R_{20} =6$ $\mu $m.}
\label{fig4}
\end{figure}

\begin{figure} 
\includegraphics[width=7.8cm]{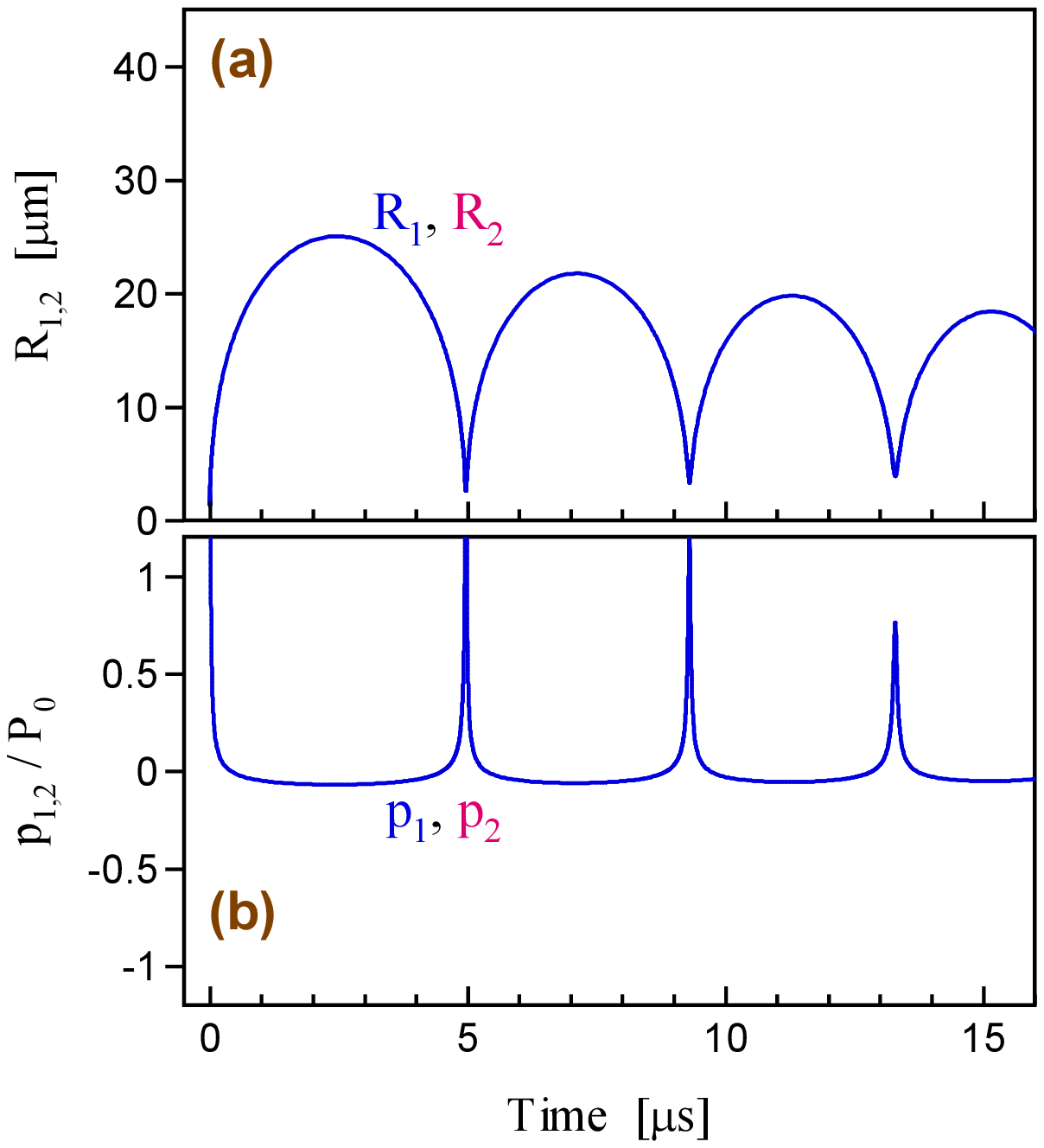}
\caption{(Color online) Same as Fig.~\ref{fig2}, but for $R_{10} =R_{20} =10$ $\mu $m.}
\label{fig5}
\end{figure}

This difficulty is resolved by considering bubble-bubble interaction. As 
well known, when a pressure wave propagating in water hits the water-air 
interface, most of its energy is reflected and a reflected wave is produced 
whose phase is opposite to the incident wave. This is because air's acoustic 
impedance is much smaller than water's. It is also known that negative 
pressure waves generated when strong positive pulses from collapsing bubbles 
hit a water-air interface can be strong enough to cause secondary cavitation 
\cite{refRefadd10}. Since the surface of gas bubbles considered here is also a free 
surface, it would be able to produce negative pulses. In Fig.~\ref{fig2}, we show a 
result for a double-bubble case. Here, we used the same parameters as in the 
above example except for $D_{12} =10R_{10} $. In this case, the change of 
oscillation amplitude is non-monotonic because of the modulation effect due 
to bubble-bubble interaction. In the bubble emitted pressures presented in 
Fig.~\ref{fig2}(b), in contrast to the single-bubble case, not only positive pulses 
but also negative pulses appear. In order to confirm the robustness of this 
observation, we performed a parametric study. The result is presented in 
Fig.~\ref{fig3}, where the maximum amplitudes of the negative pulses for $R_{10} 
=10$ $\mu $m and $D_{12} =10R_{10} $ are shown as functions of $R_{20} $. 
This proves that strong negative pulses are emitted in most cases. From this 
figure it is also found that the maximum amplitudes have a complicated 
dependence on the ambient radius. The radius-time and pressure-time curves 
for $R_{20} =6$ $\mu $m and $R_{20} =10$ $\mu $m are shown 
in Figs.~\ref{fig4} and \ref{fig5}, 
respectively. For $R_{20} =6$ $\mu $m, negative pulses are found but their 
amplitudes are very low. For $R_{20} =10$ $\mu $m (i.e., $R_{20} =R_{10} )$, 
no negative pulses are found, implying that systems of identical bubbles do 
not emit negative pulses.

\begin{figure} 
\includegraphics[width=8.4cm]{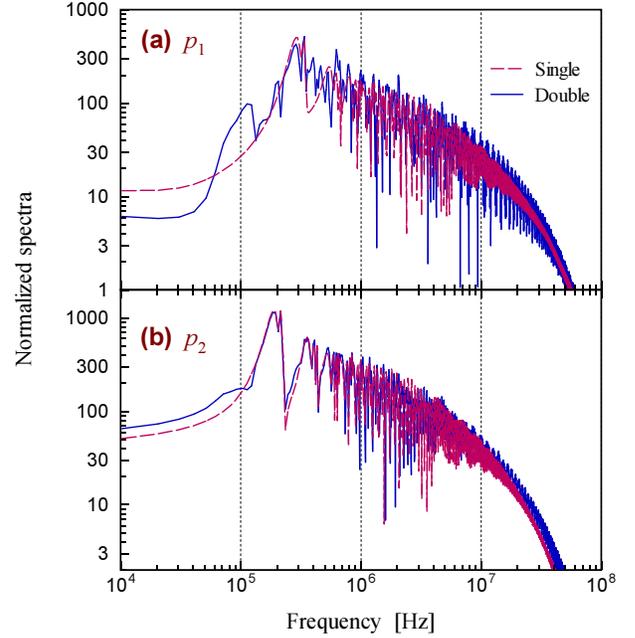}
\caption{(Color online) Frequency spectra of $p_1 /P_0 $ (a) and $p_2 /P_0 $ 
(b) for $R_{10} =10$ $\mu $m and $R_{20} =15$ $\mu $m. The dashed and solid 
curves are for the single- and double-bubble cases, respectively. The 
spectra were generated using the data shown in Figs.~\ref{fig1}(b) and \ref{fig2}(b) in a time 
period from $t=2.5$ $\mu $s to $100$ $\mu $s. The sampling frequency assumed 
is 500 MHz.}
\label{fig6}
\end{figure}

\begin{figure} 
\includegraphics[width=8.4cm]{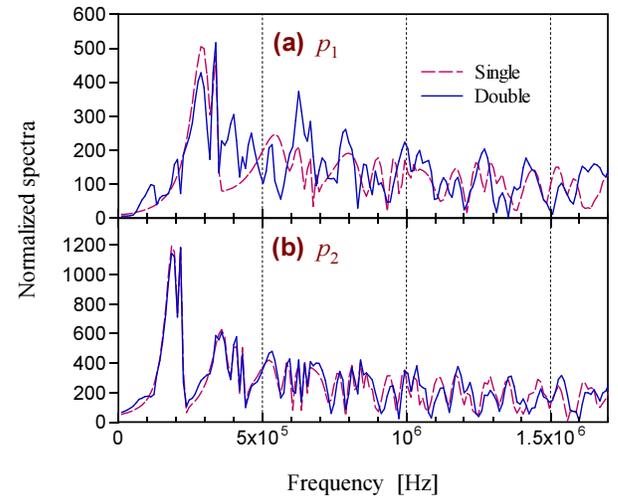}
\caption{(Color online) Frequency spectra of $p_1 /P_0 $ (a) and $p_2 /P_0 $ 
(b) for $R_{10} =10$ $\mu $m and $R_{20} =15$ $\mu $m in a frequency range 
around the peak frequencies.}
\label{fig7}
\end{figure}

\begin{figure} 
\includegraphics[width=8.4cm]{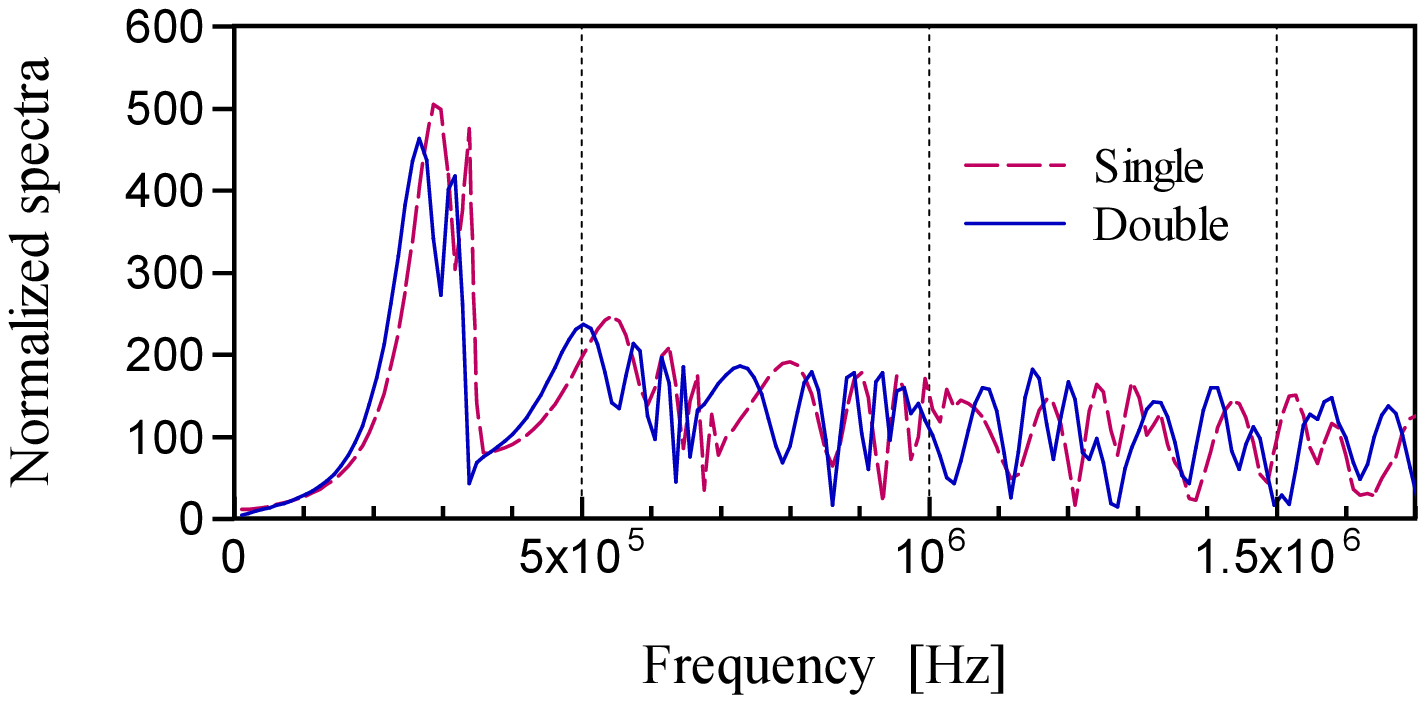}
\caption{(Color online) Same as Fig.~\ref{fig7}, but for 
$R_{10} =R_{20} =10$ $\mu $m.}
\label{fig8}
\end{figure}

The frequency spectra for $R_{10} =10$ $\mu $m and $R_{20} =15$ $\mu $m taken 
from $p_1 /P_0 $ and $p_2 /P_0 $ in Figs.~\ref{fig1}(b) and \ref{fig2}(b) 
are shown in Fig.~\ref{fig6}. 
All of the presented spectra show a broad distribution. The peak frequencies 
in the single-bubble case are 0.287 MHz for $p_1 $ and 0.185 MHz for $p_2 $, 
which are roughly the same as the repetition rates of the positive pulses 
deduced from Fig.~\ref{fig1}(b) (about 0.27 MHz and 0.17 MHz, respectively). We found 
several qualitative differences between the spectra in the single- and 
double-bubble cases. One of the differences is 
clearly seen in Fig.~\ref{fig7}, which 
shows the spectra in a frequency range around the peak frequencies. The 
spectrum of $p_1 $ in the double-bubble case has a more complex structure 
than that in the single-bubble case, particularly in a frequency range 
between 0.1 MHz and 1 MHz: A number of characteristic peaks are added and 
the spectrum structure becomes much denser through the bubble-bubble 
interaction. Compared to $p_1 $, the change in the spectrum of $p_2 $ is 
small: No qualitative differences are found between $p_2 $ in the single- 
and double-bubble cases. This suggests that the larger bubble (bubble 2) has 
a more significant influence on the neighboring bubble and the dynamics of 
the neighboring smaller bubble (bubble 1) is thus changed more drastically 
(the same tendency is found also in systems of linearly oscillating bubbles 
in a sinusoidal sound field \cite{ref6}). When $R_{10} =R_{20} =10$ $\mu $m, the 
difference between the spectra in the single- and double-bubble cases is 
very small, as can be expected from the above observation that only positive 
pulses are emitted for $R_{10} =R_{20} $; see Fig.~\ref{fig8}. The peak frequency is 
decreased by bubble-bubble interaction, which appears to be consistent with 
the fact that the natural frequency of identical bubbles oscillating in 
phase each other is lower than that of the individual bubbles 
\cite{ref7,ref8,refRefadd11}.

\begin{figure} 
\includegraphics[width=7.8cm]{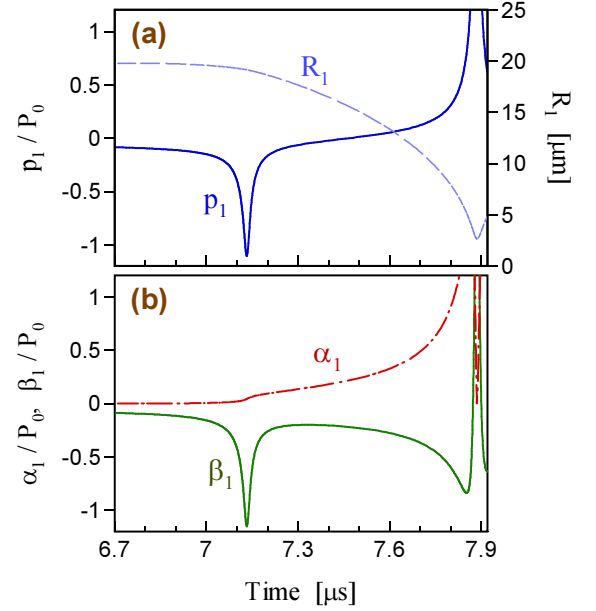}
\caption{(Color online) (a) Close-up view of $p_1 $ and $R_1 $ 
in Fig.~\ref{fig2} around the 
time of the first negative pulse from bubble 1, and (b) inertia ($\alpha 
_1 )$ and acceleration ($\beta _1 )$ portions of $p_1 $ in the same period. 
$p_1 $, $\alpha _1 $, and $\beta _1 $ are normalized by $P_0 $.}
\label{fig9}
\end{figure}

Let us consider how the negative pulses were produced. As can be seen in 
Fig.~\ref{fig2}(b), the negative pulses from a bubble coincide with the positive 
pulses emitted by the other bubble at its collapse. This observation tells 
us that the negative pulses are produced when the positive pulses hit the 
surface of the neighboring bubble. Shown in Fig.~\ref{fig9} is a close-up view of 
$p_1 $ and $R_1 $ in Fig.~\ref{fig2} around the time ($t\approx 7$ $\mu $s) of the 
first negative pulse from bubble 1. The figure also shows the inertia and 
acceleration portions of $p_1 $, that is,
\begin{equation}
\label{eq5}
\alpha _1 =\frac{2\rho R_1 \dot {R}_1^2 }{r_1 }
\end{equation}
and
\begin{equation}
\label{eq6}
\beta _1 =\frac{\rho R_1^2 \ddot {R}_1 }{r_1 }.
\end{equation}
One can see that the surface of bubble 1 is strongly decelerated in the 
period when it emits the negative pulse, though no noticeable disturbance is 
seen in $R_1 $ because the duration of the deceleration is very short. This 
strong deceleration is clearly caused by the strong positive pulse from 
bubble 2, which impulsively compresses bubble 1. Deceleration is also 
observed in a following period where the bubble is shrinking, but no 
negative pressure is found because a large inertia ($\alpha _1 )$ cancels 
the effect from the deceleration ($\beta _1 )$.

In order to more deeply understand the mechanism of negative pulse 
generation, we perform theoretical studies based on the coupled 
Keller-Miksis equations. First, we consider the relation between the 
amplitudes of the incident and reflected pulses. If $R_1 \gtrsim R_{10}$ 
(i.e., $p_{b,1} \lesssim P_0$) and $\dot {R}_1 \ll c$ at the time of 
collapse of bubble 2, where $\ddot {R}_2 $ has a very large value, only the 
acceleration terms in Eq.~(\ref{eq1}) may be significant for bubble 1 and the 
remaining terms are thus negligible. 
Hence, the equation and also Eq.~(\ref{eq4}) 
are reduced to
\begin{equation}
\label{eq7}
\left( {1+\frac{1}{A_R }} \right)R_1 \ddot {R}_1 =-\frac{1}{D_{12} }R_2^2 
\ddot {R}_2 ,
\end{equation}
\begin{equation}
\label{eq8}
p_i =\frac{\rho }{r_i }R_i^2 \ddot {R}_i \quad \mbox{for }i=1,2.
\end{equation}
In Eq.~(\ref{eq7}), $A_R $ is an acoustic Reynolds number defined here as
\begin{equation}
\label{eq9}
A_R \equiv \frac{\rho cR_1 }{4\mu }.
\end{equation}
Since $A_R $ is much larger than unity ($A_R \simeq 973$ for 
$R_1 =2.6$ $\mu $m, a typical minimum radius at collapse), 
Eq.~(\ref{eq7}) is further reduced to
\begin{equation}
\label{eq10}
R_1 \ddot {R}_1 =-\frac{1}{D_{12} }R_2^2 \ddot {R}_2 ,
\end{equation}
If one measures $p_1 $ and $p_2 $ at the same distance from the 
corresponding bubbles (i.e., $r_1 =r_2 )$, Eq.~(\ref{eq10}) is rewritten 
using Eq.~(\ref{eq8}) as
\begin{equation}
\label{eq11}
p_1 =-\frac{R_1 }{D_{12} }p_2 ,
\end{equation}
Equation (\ref{eq10}) proves that $\ddot {R}_1 $ has the opposite sign from that of 
$\ddot {R}_2 $, that is, the surface of bubble 1 is strongly decelerated at 
the moment when bubble 2 collapses, where the surface of bubble 2 is 
strongly accelerated due to its high internal pressure. From Eq.~(\ref{eq11}), one 
finds that the ratio between $p_1 $ and $p_2 $ is simply determined by $-R_1 
/D_{12} $, and that the amplitude of the negative reflected pulse is large 
if $R_1 $ is large when the pressure pulse from bubble 2 hits bubble 1. The 
numerical result shown in Fig.~\ref{fig2}(b) gives $p_1 /p_2 =-0.200$ and 
$-R_1 /D_{12} =-0.192$ for the first negative pulse from bubble 1 
(at $t\approx 7$ $\mu $s), and $p_2 /p_1 =-0.243$ and $-R_2 /D_{12} =-0.231$ 
for the second 
negative pulse from bubble 2 ($t\approx 8$ $\mu $s), both of which reveals a 
reasonable agreement. However, Eq.~(\ref{eq11}) gives a less accurate result for the 
first negative pulse from bubble 2 ($t\approx 5$ $\mu $s): $p_2 /p_1 =-0.507$ 
but $-R_2 /D_{12} =-0.369$. This may be because $p_1 $ and $p_2 $ are not 
large enough to fully satisfy the assumptions used in deriving Eq.~(\ref{eq10}).

Next, we consider what occurs when $R_{10} =R_{20} $. 
As shown in Fig.~\ref{fig5}, no 
negative pulses are emitted for $R_{10} =R_{20} =10$ $\mu $m, although strong 
positive pulses are emitted which definitely hit the neighboring bubble. 
Here we attempt to explain this observation. For $R_{10} =R_{20} $ and $R_1 
=R_2 $, at the final stage of bubble collapse where $\ddot {R}_1 $ and 
$p_{b,1} $ are very large and $\dot {R}_1 \approx 0$, Eq.~(\ref{eq1}) may be reduced 
to
\begin{equation}
\label{eq12}
\rho R_1 \ddot {R}_1 =p_{b,1} -\rho \left( {\frac{1}{A_R }+\frac{R_1 
}{D_{12} }} \right)R_1 \ddot {R}_1 
\end{equation}
where we neglected the surface-tension and viscous forces and $P_0 $ since 
their magnitudes should be much smaller than the internal pressure $p_{b,1} 
$. Since $1/A_R $ ($\simeq 1/977$ for $R_1 =2.61$ $\mu $m at the first 
collapse) is much smaller than $R_1 /D_{12} $ ($\simeq 1/38)$, Eq.~(\ref{eq12}) is 
further reduced to
\begin{equation}
\label{eq13}
\rho R_1 \ddot {R}_1 =p_{b,1} -\frac{\rho }{D_{12} }R_1^2 \ddot {R}_1 .
\end{equation}
From this we have
\begin{equation}
\label{eq14}
\ddot {R}_1 =\frac{p_{b,1} }{\displaystyle{ \rho R_1 \left( {1+\frac{R_1 }{D_{12} }} \right) }}.
\end{equation}
This says that $\ddot {R}_1 $ is positive at the bubble collapse and hence 
the bubbles emit positive pulses. This conclusion is consistent with the 
above numerical finding and is not altered even if $1/A_R $ is considered.

This result comes from the fact that, in the considered case where the 
bubbles collapse at the same time, the sound pressure $\frac{\rho }{D_{12} 
}R_1^2 \ddot {R}_1 $ from the neighboring bubble cannot exceed the bubbles' 
internal pressure $p_{b,1} $. The sound pressure generated by bubble 2, 
whose internal pressure is $p_{b,1} $, must be smaller than $p_{b,1} $ when 
it is measured at the position of bubble 1, because the bubbles are 
separated by a finite distance $D_{12} $. Since the internal pressure of 
bubble 1 is also $p_{b,1} $, it is not exceeded by the sound pressure, and 
the right-hand side of Eq.~(\ref{eq13}) is thus positive: in other words, bubble 1 
emits a strong positive pulse, resulting from the high internal pressure, 
whose absolute amplitude is greater than that of the negative reflected 
pulse. This finding suggests that the bubble surface, a free surface, does 
not always produce negative reflected pulses and the characteristics of the 
reflected waves depend on the state of the gas in the bubbles. The negative 
pulses found in Figs.~\ref{fig2}(b) and \ref{fig4}(b) are produced because the positive pulses 
hit the neighboring bubble when its internal pressure is not high.

\begin{figure} 
\includegraphics[width=7.8cm]{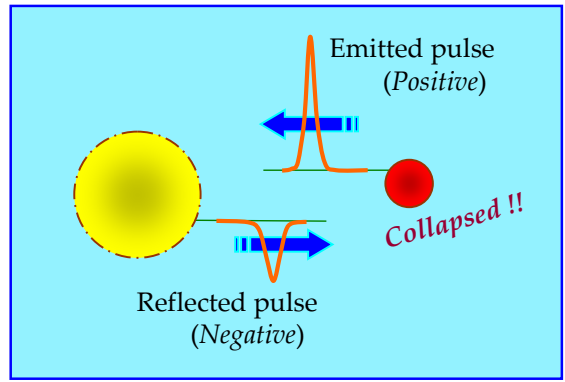}
\caption{(Color online) Noise generation process. Positive pressure pulses 
are emitted by collapsed bubbles (the right circle) and propagate through 
the surrounding liquid. The positive pulses generate negative reflected 
pulses when they hit and impulsively compress neighboring bubbles (the left 
circle). The positive and negative pulses will make up a cavitation noise. 
Note that the pulse profiles shown represent those measured at a fixed 
location, as functions of time. The wavelengths of the pulses are much 
larger than the bubble radii.}
\label{fig10}
\end{figure}

\section{Summary}
\label{secIV}

We have studied the interaction of microbubbles through pressure pulses to 
suggest a possible origin of the negative pulses found 
in broadband cavitation noise. 
The proposed scenario of noise generation is summarized as follows: When a 
large cavitation bubble collapses and fragments, a number of microbubbles 
are formed \cite{ref1}. The microbubbles expand rapidly and then collapse to emit 
positive pressure pulses (the right half of Fig.~\ref{fig10}). The positive pulses 
hit and impulsively compress neighboring bubbles to cause a brief but strong 
deceleration of the bubble surface. This deceleration, which creates a 
strong tension in the surrounding liquid, produces negative reflected pulses 
(the left half of Fig.~\ref{fig10}), and then a signal consisting of positive and 
negative pulses is generated. If the time-delay effect is considered, the 
positive and associated negative pulses are measured with a time interval 
determined by the relative position of the bubbles and the sound speed or 
shock-wave velocity of water (which can highly increase in the vicinity of 
collapsing bubbles \cite{refRefadd2,refRefadd5}). Spectral analysis has revealed that 
the frequency spectrum of the cavitation noise, particularly from the smaller 
bubble in a double-bubble system, becomes much more complicated and denser 
by bubble-bubble interaction. Though only a few negative pulses were 
observed in the double-bubble cases studied here, considering a larger 
number of bubbles and multiple scattering of sound may allow us to explain 
the large number of pressure pulses found in the recorded broadband noise. 

Discussing further details of the noise generation process, we have revealed 
that the amplitudes of the negative reflected pulses depend on the 
instantaneous radius and the state of the internal gas of the reflecting 
bubble. Interestingly, no negative pulses are generated when a system of 
identical bubbles is considered. This is because the positive pulse from a 
bubble hits the neighboring bubble just when it collapses, at which moment 
the bubble's internal pressure is higher than the pressure of the incident 
positive pulse. This observation suggests that the surface of gas bubbles, a 
free surface, does not always produce a negative reflected pulse, and also 
that negative pulses cannot be described by theoretical models that consider 
only systems of identical bubbles. The presented findings would be useful to 
understand not only the shrimp emitted acoustic signal but also other types 
of cavitation noise found, e.g., in fluid machinery \cite{ref18} and medical 
applications \cite{ref19}.

\acknowledgments
{
This work was partly supported by the Ministry of Education, Culture, 
Sports, Science, and Technology of Japan through a Grant-in-Aid for Young 
Scientists (B) (No.~20760122) and by the Japan Society for the Promotion of 
Science through a Grant-in-Aid for Scientific Research (B) (No.~20360090).
}


\begin{thebibliography}{}
\bibitem{ref16} Similar acoustic signals from cavitation bubbles can be 
found also in different fields; see, e.g., Ref.~\cite{ref18}.
\end{thebibliography}
\end{document}